\documentstyle[]{mn}

\newif\ifAMStwofonts
\input{epsf}
\newcommand{\Halpha}{H$\alpha$}
\newcommand{\VR}{V$_{\rm rad}$}
\newcommand{\Reff}{R$_{\rm e}$}

\newcommand{\degrees}{$^{\circ}$}
\newcommand{\ergcms}{erg.cm$^{-2}$s$^{-1}$}
\newcommand{\kms}{~km~s$^{-1}$}
\newcommand{\OIII}{[O~{\small III}]}
\ifoldfss
  \ifCUPmtlplainloaded \else
    \NewTextAlphabet{textbfit} {cmbxti10} {}
    \NewTextAlphabet{textbfss} {cmssbx10} {}
    \NewMathAlphabet{mathbfit} {cmbxti10} {} % for math mode
    \NewMathAlphabet{mathbfss} {cmssbx10} {} %  "   "    "
  \fi
  \ifAMStwofonts
    \ifCUPmtlplainloaded \else
      \NewSymbolFont{upmath} {eurm10}
      \NewSymbolFont{AMSa} {msam10}
      \NewMathSymbol{\upi}     {0}{upmath}{19}
      \NewMathSymbol{\umu}     {0}{upmath}{16}
      \NewMathSymbol{\upartial}{0}{upmath}{40}
      \NewMathSymbol{\leqslant}{3}{AMSa}{36}
      \NewMathSymbol{\geqslant}{3}{AMSa}{3E}

    \fi
  \fi
\fi % End of OFSS

\ifnfssone
  \newmathalphabet{\mathit}
  \addtoversion{normal}{\mathit}{cmr}{m}{it}
  \addtoversion{bold}{\mathit}{cmr}{bx}{it}
  \newmathalphabet{\mathbfit} % math mode version of \textbfit{..}
  \addtoversion{normal}{\mathbfit}{cmr}{bx}{it}
  \addtoversion{bold}{\mathbfit}{cmr}{bx}{it}
  \newmathalphabet{\mathbfss} % math mode version of \textbfss{..}
  \addtoversion{normal}{\mathbfss}{cmss}{bx}{n}
  \addtoversion{bold}{\mathbfss}{cmss}{bx}{n}
  \ifAMStwofonts
    \ifCUPmtlplainloaded \else
      \UseAMStwoboldmath
      \makeatletter
      \new@mathgroup\upmath@group
      \define@mathgroup\mv@normal\upmath@group{eur}{m}{n}
      \define@mathgroup\mv@bold\upmath@group{eur}{b}{n}
      \edef\UPM{\hexnumber\upmath@group}
      \new@mathgroup\amsa@group
      \define@mathgroup\mv@normal\amsa@group{msa}{m}{n}
      \define@mathgroup\mv@bold\amsa@group{msa}{m}{n}
      \edef\AMSa{\hexnumber\amsa@group}
      \makeatother
      \mathchardef\upi="0\UPM19
      \mathchardef\umu="0\UPM16
      \mathchardef\upartial="0\UPM40
      \mathchardef\leqslant="3\AMSa36
      \mathchardef\geqslant="3\AMSa3E
    \fi
  \fi
\fi % End of NFSS release 1

\ifnfsstwo
  \DeclareMathAlphabet{\mathbfit}{OT1}{cmr}{bx}{it}
  \SetMathAlphabet\mathbfit{bold}{OT1}{cmr}{bx}{it}
  \DeclareMathAlphabet{\mathbfss}{OT1}{cmss}{bx}{n}
  \SetMathAlphabet\mathbfss{bold}{OT1}{cmss}{bx}{n}
  \ifAMStwofonts
    \ifCUPmtlplainloaded \else
      \DeclareSymbolFont{UPM}{U}{eur}{m}{n}
      \SetSymbolFont{UPM}{bold}{U}{eur}{b}{n}
      \DeclareSymbolFont{AMSa}{U}{msa}{m}{n}
      \DeclareMathSymbol{\upi}{0}{UPM}{"19}
      \DeclareMathSymbol{\umu}{0}{UPM}{"16}
      \DeclareMathSymbol{\upartial}{0}{UPM}{"40}
      \DeclareMathSymbol{\leqslant}{3}{AMSa}{"36}
      \DeclareMathSymbol{\geqslant}{3}{AMSa}{"3E}
    \fi
  \fi
\fi % End of NFSS release 2

\ifCUPmtlplainloaded \else
  \ifAMStwofonts \else % If no AMS fonts
    \def\upi{\pi}
    \def\umu{\mu}
    \def\upartial{\partial}
  \fi
\fi

\title{Galaxy kinematics from counter-dispersed imaging}
\author[N. G. Douglas and K. Taylor]
       {N.~G.~Douglas,$^1$
	K.~Taylor,$^2$\\
        $^1$Kapteyn Astronomical Observatory, Groningen, Netherlands\\
        $^2$Anglo-Australian Observatory, Epping, NSW, Australia} 
\date{in original form 1998 April 10}

\pagerange{\pageref{firstpage}--\pageref{lastpage}}
\pubyear{1998}

\begin{document}

\maketitle

\label{firstpage}

\begin{abstract}
Determining the internal kinematics of a galaxy from planetary nebulae
(PN) is usually a two step process in which the candidate PN 
are first identified
in a target galaxy and then, in a follow up run, spectra
are obtained. We have implemented a new technique in which
two dispersed images at the wavelength of the \OIII\ emission line
at 5007\AA\ 
are combined to yield positions, magnitudes and velocities of the
PN population in a single step. 
A reduction in observing time of about a factor 2 
is attainable. We present here the proof-of-principle results.

\end{abstract}


\section[]{Introduction}

The dynamics of distant galaxies are often studied by measuring the
radial velocity field of the integrated stellar light (e.g.  optical
observations using long-slit or Fabry-Perot) or of the associated gas
(e.g.  HI observations).  This data is compared with models in order to
derive information on structure and dynamics, or inverted to arrive at a
three-dimensional model directly.  Integrated light techniques, whether
using absorption lines or emission lines, tend to be limited by surface
brightness to about 1 effective radius (\Reff) from the nucleus.  HI
observations remain the most sensitive means of studying late-type
galaxies but in general are not applicable to early-type (E - S0)
galaxies owing to their paucity of gas.  Moreover HI may not trace the
same kinematics as the stellar component.  The same remark applies to
globular clusters -- which for example while orbiting in the same
gravitational potential as the stars may not, as a population, be
rotating, while the galaxy as a whole is \cite{Arna94}.
Measurements at large radii are important for numerous reasons.  For
example some galaxies exhibit, with respect to large-scale kinematics,
counter-rotating or differentially-rotating cores.  Kinematical
conclusions based on only circumnuclear studies can be misleading: in
NGC~1399 the rotational velocity falls to a minimum at about 20~arcsec
from the centre \cite{Dono95} but on a larger
scale the rotation is still increasing out to 200~arcsec \cite{Arna94} and a
considerable fraction of the angular momentum is located outside 1\Reff. 

The use of PN as radial velocity probes enables measurements over a
larger extent, typically $\sim 5$\Reff, in all galaxy types, and has a
long and successful history.  However, the usual procedure is to measure
the flux and position of each PN by direct imaging and its radial
velocity by the use of fibre spectroscopy.  This is a time-consuimg
sequence of operations and is prone to failure. 

We propose, and have tested, a simpler technique called
counter-dispersed imaging (CDI) in which the required information is
obtained with just two narrow-band images.  After reviewing the status
quo we describe CDI in greater detail, including issues such as overal
efficiency and the steps required for calibration.  We present the
results of proof-of-principle observations conducted some years ago (AAT)
and make reference to  more recent results (WHT).  Mention is made
of a dedicated instrument under development, but this paper is primarily
intended to illustrate the use of the technique with existing
instrumentation. 

\section[]{PN as tracers}

\subsection{Characteristics of PN populations}
\label{cPN}

The PN appears to be a phase in the life cycle of all medium-mass
stars and, independent of type or metallicity,  no galaxy deficient in PN
has yet been found. Most of the visible emission appears in a single line
(\OIII\ at a rest wavelength of 5006.875\AA).  PN brightness in
this line is commonly
given in  magnitudes according to:
\begin{equation}
m_{5007} = -2.5 \log F_{5007}  - 13.74
\end{equation}
where $ F_{5007}$ is in \ergcms\ and the constant is such that a star
of the same magnitude would appear to be equally bright in a V-band filter.
We will drop the numerical subscript.
On theoretical grounds one expects the luminosity
function $\Phi$ to follow an exponential increase with increasing 
magnitude (Hui et al.\ 1993b eqn~2) but observations suggest a sharp cutoff
at the bright end. Accordingly Ciardullo et al.\ (1989)
introduced an empirical modification:
\begin{equation}
\Phi(m) \propto e^{0.307m}[1-e^{3(m*-m)}] 
\label{pnlf}
\end{equation}
where the parameter m* defines the bright cutoff. 
For M31, Ciardullo et al.\ give the
bright-end of the PNLF cutoff as M* = -4.48.
The luminosity specific
density of PN is not constant for all galaxies but does not differ greatly
from the value determined by Hui et al.\ for NGC~5128 \cite{Hui93b}:

\begin{equation}
\beta_{2.5} \sim 100 \times (10^9 L_{\sun})^{-1}
\label{beta25}
\end{equation}
where reference is to the B-band luminosity. 
This is the number of PN in the top 2.5~mag of the PNLF,
e.g. a M=-21 galaxy would have would have $\sim$ 1900 such PN.

\subsection{Radial Velocity Measurements}

 The intrinsic linewidth of the \OIII\ feature being only about 21\kms,
PN are ideal probes of galaxian velocity fields, which have dispersions
much greater than this.  In practice PN radial velocities can be
determined by fibre spectroscopy to $\pm$10\kms. The high brightness
in the \OIII\ line makes the PN easy to detect.
 
 The other requirement in order to be useful as tracers is that the PN
be abundant in sufficient number.  To establish the velocity or the
velocity dispersion to 25\% accuracy requires $\sim$ 15 PN in a given
projected area.  
    Now for a $m_{\rm B} \sim 11$ galaxy at a distance of
10~Mpc Eqn~\ref{beta25} implies a total of $\sim 300$ PN in the top
2.5~mag of their brightness distribution.  In a typical observation
  % TO DO ? (discussed later) 
about twice this number could be expected to be
detected.  If we define spatial bins to be 60\degrees\ in azimuth and 1
scale length (1/e in surface brightness) in radius then the number of PN
per bin is around 20 at r=1\Reff\ and 10 at r=4\Reff. 
 Although the projected density of PN falls in direct proportion to the
surface brightness of the galaxy, the increasing size of the spatial
bins partially compensates.  Furthermore, although not taken into
account here, the detection efficiency increases since the galaxy's own
background light is a major source of noise.  Thus, PN will be present
in sufficient numbers in very many situations, albeit not in galaxies of
low surface brightness.  Moreover it has been shown that important
results (such as establishing the presence of rotation) can be obtained
even with 60 PN
\cite{Arna94}.
The problem of obtaining three-dimensional velocity
models from only radial velocities
\cite{Hui93a} is of course
common to all techniques in current use.

\subsection{Distance indicators}

In addition to being a dynamical probe, the PN
are a reliable secondary distance indicator through the
`standard-candle' use of the PNLF. 
This relies on
the assumption that M* (\S\ref{cPN}) is a universal constant, which has
been tested in a series of papers by Jacoby and colleagues (the latest
being Feldmeier et al.\ 1997).  For distance determination
(and to a lesser extent kinematics) it is necessary to screen the
luminosity function for unresolved HII regions, which appear similar
to PN \cite{Feld97}.

\section[]{Traditional observing technique}

Detection of the PN in \OIII\ emission 
is usually done by imaging the field with on-band
and off-band filters \cite{Jacoby92}. While continuum sources
show up in both, PN show up only in the on-band image.
For maximum
sensitivity the on-band filter is no wider than need be to
include the range of velocities required (if the systemic velocity
is known) while  the off-band
filter can be considerably broader. Typically, therefore, 
if $T$ is the on-band
integration time the total time required is $\sim 1.3 T$.
The positions of the PN then have to be determined to sub arcsecond
precision (i.e. far more precisely than required by the science)
in order to set up a subsequent observing campaign with multi-object 
spectroscopy. Aperture plates or, more recently, robotic fibre
positioners are used. Estimating the observing time for the
spectroscopy is difficult as it depends on many factors, such as the 
respective numbers of PN detected and available fibres per observation,
but from cases in the literature it can be taken to
be $\sim 4T$ giving a total of about
$5.3 T$ telescope time spread over two semesters.

\section[]{Counter-dispersed imaging}

\subsection{Basic Idea}
The principle of obtaining well-calibrated spectra of objects
in an extended field by the use of two counter-dispersed images
is not new. As early as 1947 Fehrenbach used pairs of objective prism plates
to obtain stellar radial velocities in this way. 
At an IAU meeting in 1994   
we proposed the use of a conventional slitless
spectrograph at the 5007\AA\ \OIII\ line to obtain accurate PN radial
velocities in a galaxian field \cite{iau149}.  Rotating the
entire spectrograph (see Fig.~1)
has an analogous effect to rotating an objective
prism plate.  The PN, which appear as point
sources against the dispersed background, are displaced in opposite
directions by an amount which is very nearly linear in radial velocity. 
PN pairs, once they are found and uniquely matched up, yield relative
velocities with great accuracy. 
In addition, we showed that the pairs of images could be obtained {\em
simultaneously}, if desired, by either wavefront- or amplitude-splitting. 
At a subsequent IAU meeting in  1996 we reported plans to build a dedicated
instrument \cite{iau180} and we are now participants in
a consortium to carry this out \footnote{The PN Spectrograph project
is reported at\\  www.aao.gov.au/local/www/pns/pns.html}.
\begin{figure}
\epsfxsize=80mm
\epsfbox{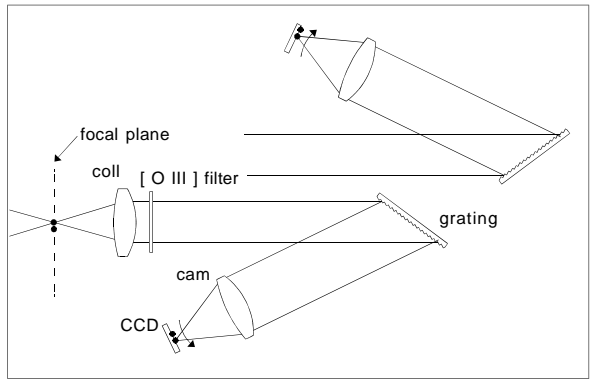}
\caption{Schematic representation of the way in which a 
slitless spectrograph
can be rotated to record, subsequently, pairs of images.
Cross correlating the images yields 
pairs of PN candidates whose separation 
is a direct measure of the radial velocity.}

\label{pnseek1}
\end{figure}

\subsection{Advantages}

A properly designed (slitless) spectrograph will have an efficiency of
about 70\% that of a direct imager due to losses at the dispersive
element.  The CDI method of obtaining PN velocities requires only two
observations, namely an on-band image at arbitrary PA and one at PA+180\degrees. 
The total integration time is therefore $\sim 2.9 T$.  This is roughly a
factor two less than that traditionally required, and the two
observations can usually be made during the same night.  The true
positions of the PN can easily be
 reconstructed from the two dispersed images, and the radial velocities
can be found - to within a constant offset - by inspection. This
velocity offset can be determined by ancillary calibration
observations (e.g. of a galactic PN) but can also be found using an
arc lamp (if a slit unit can be inserted for the purpose).
 Potentially, very accurate fluxes can be obtained since the filter
profile can be determined during the night, at the f/ ratio and
temperature appropriate to the observations and even as a function of
position on the detector, from short exposures on foreground stars. 
It is more difficult to determine the filter characteristics 
in direct imaging.

\subsection{Technical issues}

1. Almost any medium-dispersion spectrograph can be modified
for CDI although the unvignetted field may be disappointingly
small if the spectrograph is not intended for multi-object work.
Over the small ($\sim 40-60$\AA) wavelength range required the
dispersion will be found to be very constant over the field, but the
same effect as causes arc lines to be bent will cause spurious
displacements -  varying quadratically
with distance from an axis of symmetry parallel to the dispersion axis.\\
 2.  It is desirable that the dispersion not be too high because of two
undesirable effects: firstly, field is lost since PN which are dispersed
outside {\em either} image are rendered useless, and secondly,
foreground stars will produce unnecessarily large trails which
obliterate some of the field.  Suitable values of the dispersion are
around 1\AA/pixel.  If the individual PN images can be centroided to an
accuracy of 0.4 pixels then the error in the {\em differential} shift is
about 0.56 pixels or 17\kms, while for a typical bandwidth of 60\AA\ the
star trails are then an acceptable $\sim 100$ pixels in extent.\\
 3.
To a first approximation the detection threshold for direct
imaging and for slitless spectroscopy - through the same
interference filter - is the same, provided that the PSF is
similar in each case. Since the galaxian light is dispersed
in the latter case it will be found that detection close to
the nucleus will be slightly improved and away from the 
nucleus slightly worsened. \\
 4.
In the case of fibre- or slit-spectroscopy of the individual PN the
background is much reduced when compared with slitless spectroscopy or
direct imaging.  In practice a number of loss mechanisms come into play
which more than compensate for this gain, as is evident from published
results.  A comprehensive discussion of this point is beyond the scope
of this paper.

\section[]{CDI with existing instrumentation}

\subsection{Trial observations}

We modified the RGO spectrograph at the 3.9~m Anglo-Australian telescope
to establish the proof-of-principle and to test the CDI mode of
operation.  This took place in discretionary time in January 1995.  This
run is reported in more detail below. 
During a subsequent observing campaign at the William Herschel Telescope
(April 1997) the dual-beam ISIS spectrograph was modified in such a way
that it provided a dispersed \OIII\ image and, simultaneously, an
undispersed \Halpha\ image.  In addition, comparison data were taken in
CDI mode.  These observations are reported in 
Kuijken et al.\ 1998. 

\subsection{Instrument configuration}

The RGO spectrograph was used with the 82~cm focal length camera and the
1200J grating in m=2 with blaze to collimator; the plate scale was
0.236~arcsec/pixel (determined from DSS stars) and using this value and
the known grating geometry the linear dispersion was calculated to be
0.11069~\AA/pixel.  Note that no slit was used.  The \OIII\ filter was
5016/25. 
  The total efficiency during periods of good visibility
was calculated from the following estimates, based on data from
various sources:
atmosphere: 0.83, 
telescope: 0.7,
spectrograph optics:  0.42,
grating: 0.57,
filter: 0.6 and 
CCD: 0.59, giving  0.05 total efficiency.

\subsection{Observations}

\begin{table}
\caption{Log of AAT/RGO observations}
\label{AATlog}
\begin{center}
\begin{tabular}{lrrr}
\hline
&&& \\
Field             &  PA       & Date (1995)    & Exposure time\\
&&& \\
\hline
&&& \\ 
mac 2-1      	& 90         & January 25 	& 20s; 120s \\
NGC 1400  	& 90         & January 25 	& 1800s \\
NGC 1400  	& 270         & January 25 	& 2x1800s \\
NGC 1400  	& 90         & January 25 	& 1800s \\
mac 1-2       	& 90              & January 25 	& 40s \\
NGC 5128      	& 90         & January 25 	& 1800s \\
NGC 5128  	& 270         & January 25 	& 2x1800s \\
NGC 5128      	& 90         & January 25 	& 1600s; 2x900s \\
\hline
NGC 1400  	& 270         & January 26 	& 1800s \\
NGC 1400  	& 90         & January 26 	& 2x1800s \\
NGC 1400  	& 270         & January 26 	& 1500s \\
NGC 5128  	& 270         & January 26 	& 480s;1800s \\
NGC 5128  	& 90         & January 26 	& 2x1800s \\
NGC 5128  	& 270         & January 26 	& 1800s;1500s \\
NGC 5128  	& 90         & January 26 	& 1500s \\
mac 1-2       	& 90          & January 26 	& 5x40s \\
\hline
\end{tabular}
\end{center}
\end{table}

Instrument was operated as shown schematically in Fig~\ref{pnseek1}. 
Weather permitting, we attempted to observe in the sequence A-B-B-A to
minimise the effect of changes. In order to have reference data 
  we observed the well-studied galaxy NGC~5128 (field F42 of Hui et al.\
1993b) as well as NGC~1400, an E-S0 type of unknown distance.  For flux
calibration we observed the Galactic PN mac2-1 and mac1-2.  The
observing log is presented in Table~1. 
  The weather was poor during most of the run and the atmospheric
transparency variable. 
  Many of the listed exposure times are, for this reason, only nominal. 
The summed data in each of the two counter-dispersed images corresponded
to an estimated 2.0 - 2.5h observing time. 
  Several exposures were not be used and reliable photometric results
were not attainable.  Seeing varied up to 1.6". 

\subsection{Reduction}

\begin{figure*}
\label{ABframes}
%\epsfxsize=80mm
%\epsfbox{Aframe.ps}
\epsfxsize=80mm
\epsfbox{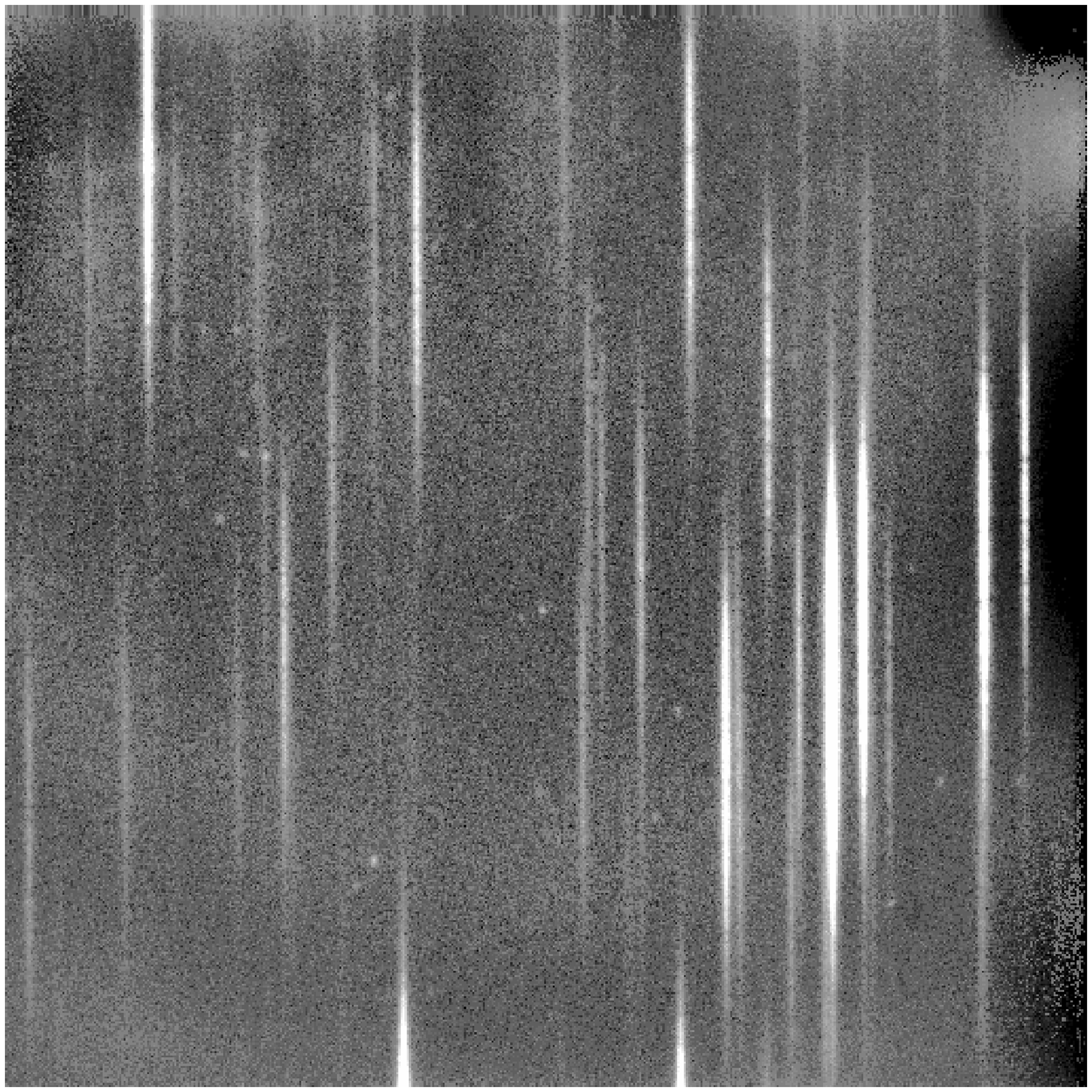}
\hspace{5mm}
\epsfxsize=80mm
\epsfbox{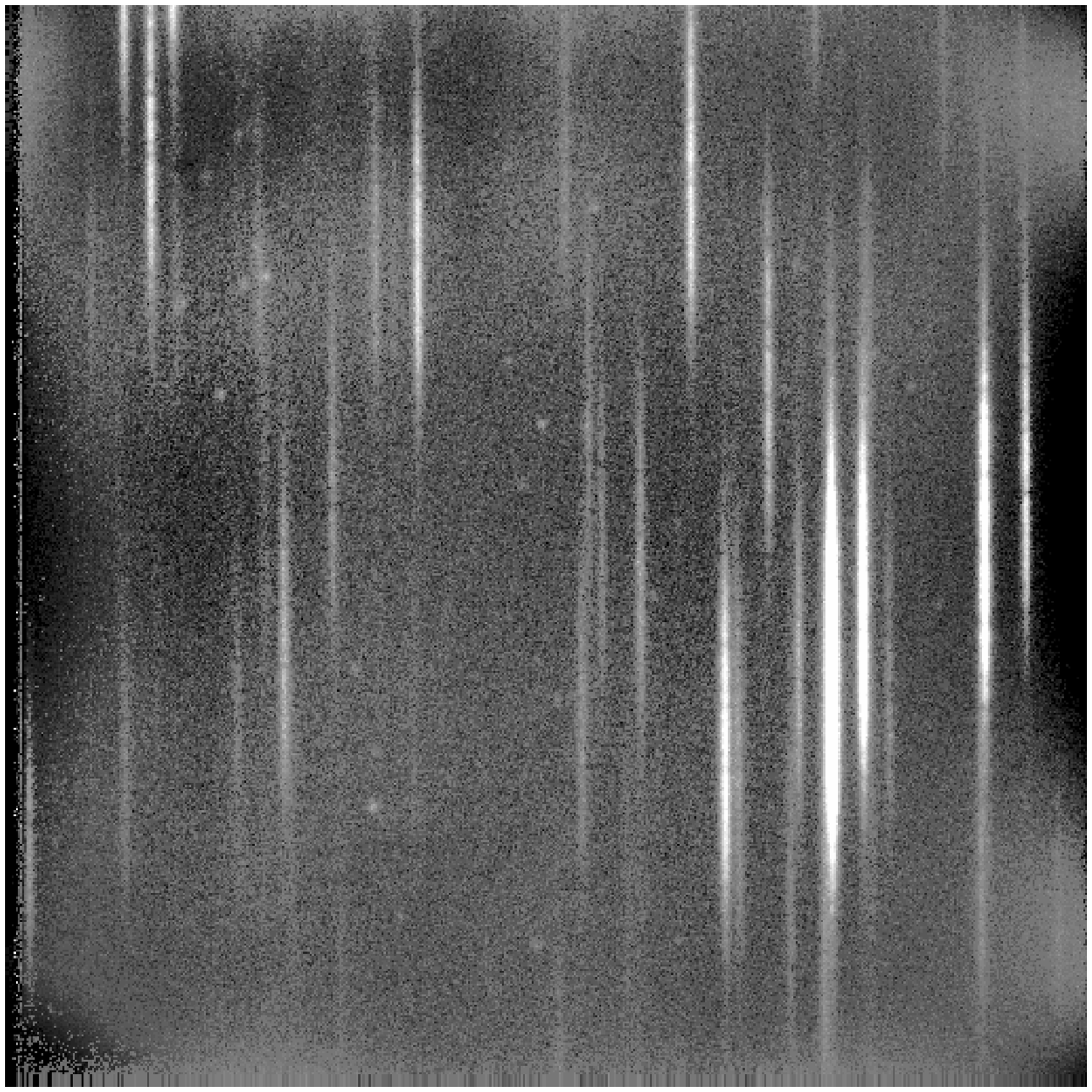}
\caption{Dispersed image of the field NGC~5128/F42. A few
PN are visible as point-like objects.}
\end{figure*}

\begin{figure*}
\label{dss}
\epsfxsize=80mm
\epsfbox{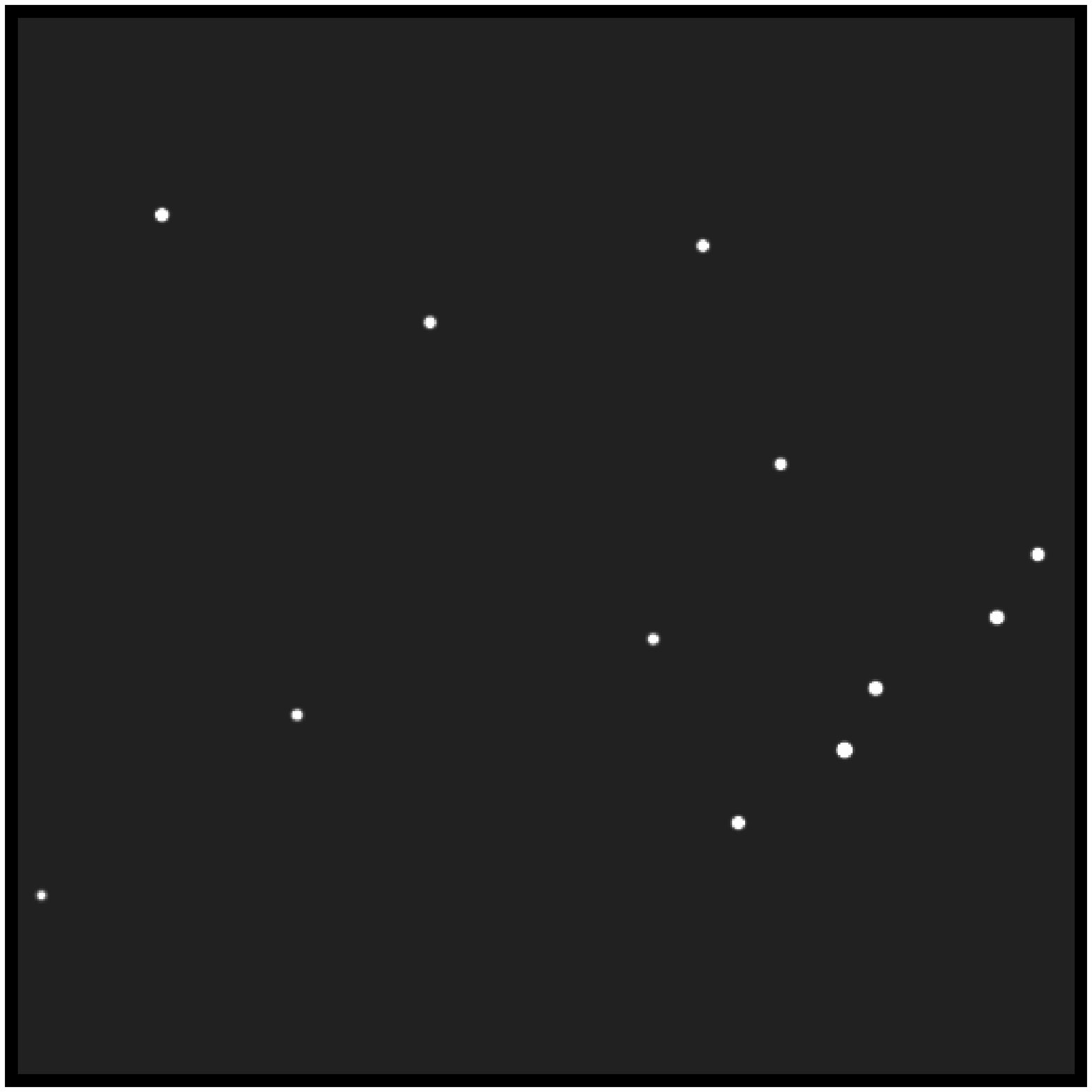}
\hspace{5mm}
\epsfxsize=80mm
\epsfbox{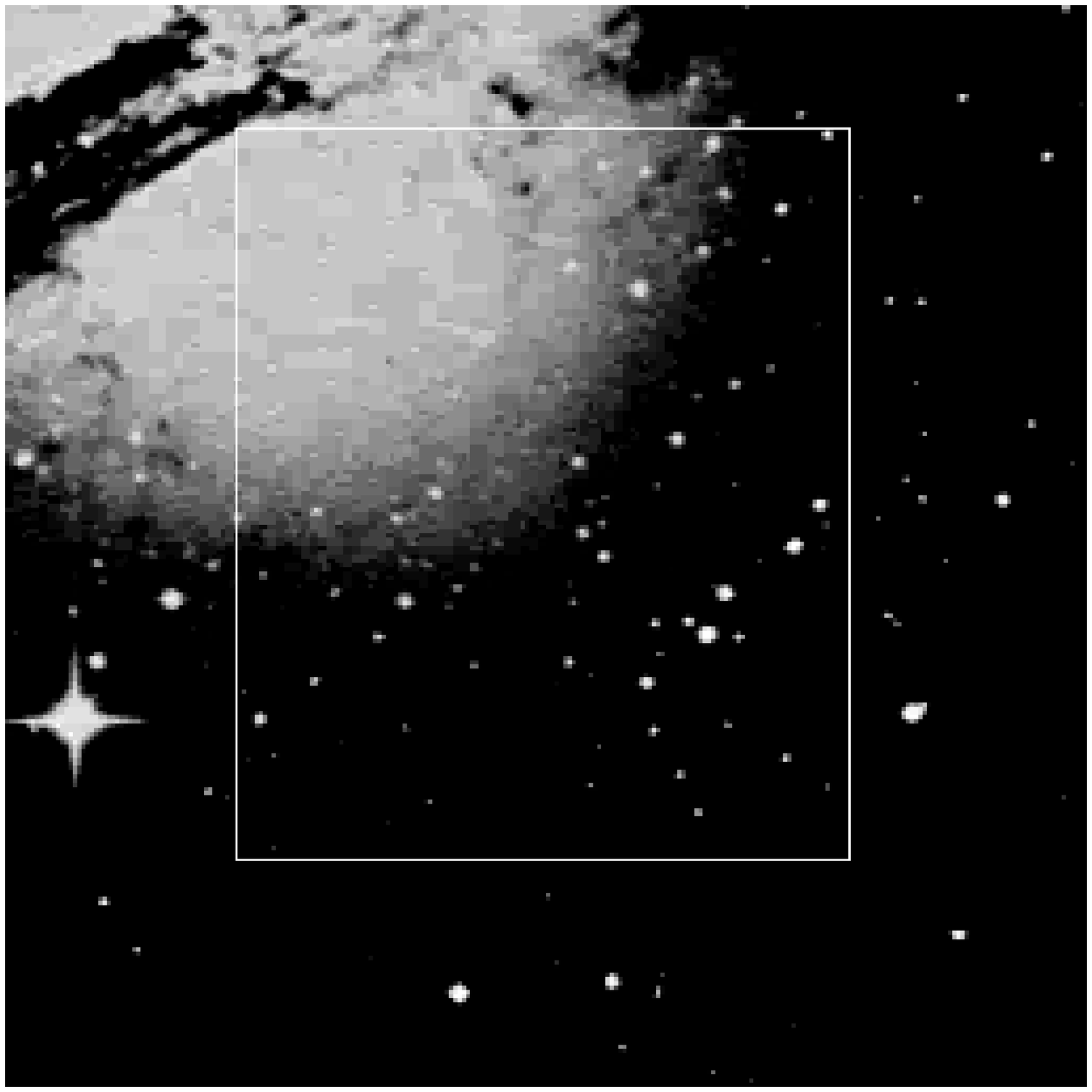}

%\epsfbox{F42.ps}

\caption{Left:
Partial stellar field reconstructed from the data in
Fig.~2 using absorption lines.
Right:
Digitised Sky Survey image (N at top) with box showing the
approximate location of our field.
The reconstructed field is easily
recognised.}

\end{figure*}

The data were reduced in IRAF using standard procedures
for bias subtraction and pixel-to-pixel variation. The strong 
vignetting pattern of the spectrograph was estimated from 
sky flats and an appropriate correction was applied - note
that this is only approximate since the final position of
a PN in the dispersed image is a function both of position in
the field {\em and} wavelength. No attempt was made to improve the
photometric correction after the `true' positions were 
reconstructed since we were not interested in high photometric accuracy.
Two dimensional, 6th order  Legendre polynomial fits were made
to the background intensity which included sky and galaxy contributions,
 and subtracted. The individual images at each PA were shifted into
registration by centroiding stellar absorption lines. This involves spectral
fitting in the dispersion direction (Y in Fig.~2)
and gaussian fitting of
the continuum intensity in the perpendicular direction (X in Fig.~2).
 The frames were then averaged using weights based on signal-to-noise ratio
and using {\tt imcombine.crrej} to reject
cosmic ray events. Simple median filtering was problematical 
because of the strong variation in signal caused by weather.
For simplicity of analysis one of the two final images was rotated
by 180\degrees.

Image registration was made difficult in the case of
NGC~1400 because the field
contains only two stars, both lacking in spectral features.
Individual PN were not visible on single
frames (for either galaxy) so these could not be used as 
markers. A crude estimate of the required shift was made by
fitting the entire dispersed image of the two stars and 
a small range around this nominal shift was then explored  
for signal.

 The co-added CDI frames  for NGC~5128/F42 are shown in
Fig.~2. To establish astrometric coordinates
a pseudo-image was created in which a stellar object 
represented the mean position of a star trail between
the two frames. This could be done very accurately
by fitting absorption lines. For each star trail three
lines were used, the mean position of
each line giving one estimate. The reconstructed (partial) stellar image 
shown in Fig.~3. Note that the relative positions
of the stars within this `constellation' depends only on
plate scales in X and Y and is quite independent of dispersion
and of the wavelengths of the absorption features.
These stars were identified on the digitised sky survey 
(also shown in Fig.~3) and their RA and
dec used to derive the plate transformation. 
 Similarly, after locating the point sources in the counter-dispersed
images and assigning them into pairs, each corresponding to a
PN, their `true' positions were reconstructed by averaging.

\subsection{Results}

\begin{table}
\label{N5128.tbl}
\caption{PN detected in the F42 region of NGC~5128. Our positions and
velocities are compared with those of Hui et al.\ (1993c, 1995). 
 }
\begin{center}
\begin{tabular}{l|lll|lll}
\hline
&&&&&&\\
I.D.   & $\alpha$(2000) &  $\delta$(2000)  & \VR &  \arcsec ($\alpha$) & \arcsec ($\delta$) & \VR \\
       &\multicolumn{3}{|c|}{~----------~~this paper~~----------~}
       &\multicolumn{3}{|c|}{~---~~Hui et al.~~---~}\\
&&&&&&\\
\hline
&&&&&&\\
4221   & 13:25:08.93 & -43:03:53.64 & 605  & 08.90 & 54.1  & 633   \\ 
4228   & 13:25:08.34 & -43:04:52.53 & 585  & 08.55 & 58.5  & n/a  \\ 
4211   & 13:25:11.26 & -43:03:07.97 & 330  & 11.22 & 08.6  & 344   \\ 
4216   & 13:25:13.69 & -43:04:30.78 & 645  & 13.69 & 31.5  & 645  \\ 
4208   & 13:25:14.17 & -43:04:54.99 & 758  & 14.18 & 55.8  & 765  \\ 
4204   & 13:25:16.45 & -43:04:03.28 & 656  & 16.42 & 04.1  & 645   \\ 
4222   & 13:25:16.85 & -43:04:12.12 & 488  & 16.83 & 12.5  & 523  \\ 
4201   & 13:25:19.90 & -43:05:28.92 & 234  & 19.92 & 30.3  & n/a   \\ 
4213   & 13:25:20.25 & -43:05:13.63 & 736  & 20.28 & 14.6  & 732  \\ 
4202   & 13:25:22.05 & -43:03:21.06 & 619  & 21.99 & 22.1  & n/a  \\ 
4206   & 13:25:22.50 & -43:03:21.82 & 584  & 22.44 & 22.6  & 599   \\ 
4205   & 13:25:22.97 & -43:03:45.55 & 438  & 22.94 & 46.6  & 447  \\ 
4241   & 13:25:16.12 & -43:05:11.95 & 511  & 16.17 & 12.9  & 520   \\ 
4279   & 13:25:16.54 & -43:05:04.53 & 391  & 16.58 & 04.5  & n/a  \\ 
4207   & 13:25:23.27 & -43:02:50.29 & 526  & 23.19 & 51.2  & 527   \\ 
4240   & 13:25:16.29 & -43:05:59.95 & 521  & 16.33 & 61.1  & 503   \\ 
4209   & 13:25:22.62 & -43:02:44.52 & 662  & 22.52 & 45.4  & 675   \\ 
4214   & 13:25:15.38 & -43:03:03.35 & 628  & 15.29 & 04.0  & 624   \\ 
4210   & 13:25:18.59 & -43:02:20.99 & 716  & 18.50 & 21.8  & 685   \\ 
4231   & 13:25:15.29 & -43:02:48.78 & 483  & 15.20 & 49.0  & 440  \\ 
4248   & 13:25:17.18 & -43:02:51.34 & 621  & 17.10 & 51.9  & 599  \\ 
4236   & 13:25:17.11 & -43:03:41.65 & 558  & 17.07 & 42.5  & 565   \\ 
4263   & 13:25:11.49 & -43:03:50.57 & 560  & 11.44 & 51.0  & 577  \\
4255   & 13:25:18.36 & -43:03:04.43 & 695  & 18.30 & 04.8  & 664 \\
&&&&&&\\
\hline
\end{tabular}
\end{center} 
\end{table}

In  NGC~5128/F42 we detected 24 PN (i.e. at both PA). 
Our results 
are shown in Table~2. We discuss how these were 
obtained and estimate errors in the following subsections:\\
\underline{\bf Positions} After
applying the plate transformation just mentioned we arrive at the
RA and dec as shown. Also shown are the
positions as listed by \cite{Hui93c} and the ID number they
ascribed to individual PN. Our positions mostly agree to within 0.5\arcsec,
proving that adequate astrometry is obtained just by averaging the
coordinates found at the two PA. Radial velocity information is 
{\em not} required for this step,
even for the stars used to derive the plate transformation.
\\
\underline{\bf Velocities}
From the dispersion coordinate (Y$_i$, in pixels) in the counter
dispersed images the radial velocity can be derived from the formula V =
0.5*(Y$_1$-Y$_2$)*$d$*C/$\lambda$ + V$_0$.  Here C and $\lambda$ (5007\AA)
have their usual meanings, $d$ is the dispersion and V$_0$ is an offset.
These last two parameters would normally be obtained by 
calibration observations
of a spectral lamp and slit, but this was not possible with the RGO
spectrograph as the slit unit had to be 
removed for the run. Calibration on a foreground star or
galactic PN of known
velocity would have been possible but in the absence of a slit any
error arising due to pointing or to flexure would have been difficult
to determine. 
 Instead we calculated $d$
from the spectrograph properties and left
V$_0$ as a free parameter to be determined by comparison
with the reference data. Before doing this, a small but
important correction was made to eliminate the effect
of `spectral curvature', i.e. the bending of the image of a slit
due to off-axis angle ($\Gamma$) at the spectrograph grating.
In our case $\Gamma \sim$ 0.8\degrees at the edges of the CCD
(left and right in Fig.~2)
leading to a spurious shift of about 4.8~pixels in the dispersion
direction. Owing to the CDI geometry the effect is additive in (Y$_1$-Y$_2$), 
appearing as a velocity offset of $\sim$ 64\kms.
 The best fit to the results of \cite{Hui95}  required V$_0$~=~467\kms and
the best fit dispersion was $d$  = 0.1107~\AA/pixel, exactly as calculated
from first principles.  The error in our velocity determination is
determined by three factors:\\
 1.  the dispersion calibration - this is well-known for the RGO
spectrograph, is constant over the field, and is confirmed by the
agreement mentioned in the previous paragraph. \\
 2.  errors in centroiding - this is easily checked by comparing 
our astrometric determinations with those from direct imaging and
is found to be of the order
0.5". This corresponds formally to $\pm 2.3$\kms.
The r.m.s. difference between our results and the
 reference data is 19.1\kms, about twice the 1$\sigma$ error
claimed by \cite{Hui95}.
\\
 \underline{\bf Photometry} The system efficiency was 0.05, as found by
observation of the Galactic PN Mac 1-2.  However the average
transparency during our integrations was about 50\%, as estimated from
the magnitudes of the PN actually observed.  For example PN \#4204 
has a flux of 1,390 in 1800s in our data.  Hui et al.\ (1993c) ascribe it $m =
23.71$, from which we deduce an overall efficiency for our system of
0.025.

In the field centred on NGC~1400 we detected no PN in approximately the
same integration time as that spent on NGC~5128 ($\sim 3$hr).  Although
the difficulty in co-adding the integrations is a complication,
considerable effort was expended in searching for the correct shifts
(only one direction is undetermined) and therefore this is not felt to
be the cause of the non-detection.  The simplest explanation is that the
brightest PN in NGC~1400 are fainter than our detection limit of $m \sim
25.2$, putting NGC~1400 at a distance of 8.6~Mpc or greater.  This is
consistent with the result that NGC~1400 is a member of the NGC~1407
group \cite{Gould93}.

\section{Discussion}

In field F42 Hui et al.\ (1995) detected 134 PN in a  narrow-band
image with a 1hr integration (not including
the time for the off-band image) with 1\arcsec seeing. In the
 unvignetted part of the RGO field (about 45\%) we would expect to have
seen about 60 PN at the same sensitivity.  We integrated for 3hr at each
PA but with a system efficiency of 0.025 as opposed to 0.3 (estimated)
for direct imaging.  Thus, our limit was approximately 1.5 mag poorer. 
Our detection of 24 PN is therefore consistent with the published
luminosity function \cite{Hui93b}, which has approximately 0.4 of the
PN more than 1.5 magnitudes above the faint cutoff.  From this one could
infer that the detection efficiency of the (slitless) spectrograph would
be equal to that of direct imaging, given equal system efficiency. 
Actually the RGO spectrograph has such high dispersion, and the field so
many foreground stars, that approximately 20\% of the CCD is obliterated
by dispersed stellar light.  Apart from this, largely
avoidable difference, we believe the foregoing statement to be correct. 

In contrast to the 1hr identification exposure, 
several hours of multi-fibre spectroscopy
were required on each field to complete the observations
\cite{Hui95}. Although fibre-fed
spectrographs constinue to improve, this phase of the observation
still remains many time less efficient than the detection phase.
The advantage of counter-dispersed imaging is that only a second 
slitless image is required.

It has been proven that CDI is a straightforward method of detecting the
PN population of an external galaxy and of determining their positions
and radial velocities.  If required, there is no doubt that accurate
photometry can be obtained although this was not demonstrated here due
to the poor observing conditions.  A dedicated CDI instrument would
return the data required for the analysis of kinematics and for PNLF
distance determinations in a single observation and with an efficiency
$\sim 2$ times better than hitherto possible. 
 
\section*{Acknowledgments}

NGD performed part of this work while on sabbatical leave at the
Anglo-Australian Observatory and the Anglo-Australian Telescope was used
to obtain data, with thanks to R. Cannon for
discretionary time.  We made use of the NASA/IPAC Extragalactic Database
     (NED) which is operated by the Jet Propulsion Laboratory, California
     Institute of Technology, under contract with the National Aeronautics and
     Space Administration. 
Thanks to Guy  Monnet (ESO) for references to Fehrenbach's
work.

\label{lastpage}

\end{document}